\documentclass{PoS}
\usepackage{amsmath}
\usepackage{amssymb,xcolor,trimclip}
\usepackage{wrapfig}

\newcommand{\qua}{{\rm q}}
\newcommand{\glu}{{\rm g}}

\title{Jet fragmentation in a QCD medium: \\ Universal quark/gluon ration and Wave turbulence}

\ShortTitle{Jet fragmentation in a QCD medium: Universal quark/gluon ration and Wave turbulence}

        \author{Y.~Mehtar-Tani\thanks{We thank J.-P.~Blaizot, E.~Iancu, A.~Kurkela, A.~Mazeliauskas, J.~F.~Paquet, K.~Tiwonyuk and D.~Teaney for insightful discussions and collaborations on the topics discussed in this proceeding. This work was supported in part by the U.S. Department of Energy, Office of Science, Office of Nuclear Physics under Award Numbers DE-FG02-97ER41014 (SS) as well as DE-FG02-00ER41132 and DE-SC0012704 (YMT). S.S. also acknowledges support by the Deutsche Forschungsgemeinschaft (DFG, German Research Foundation) through the CRC-TR 211 'Strong-interaction matter under extreme conditions' -- project number 315477589 -- TRR 211.
}\\
        Brookhaven National Laboratory, Physics Department, Upton, NY 11973, United States\\
        E-mail: \email{mehtartani@bnl.gov}}

\author{\speaker{S.~Schlichting}\\
        Fakult\"{a}t f\"{u}r Physik, Universit\"{a}t Bielefeld, D-33615 Bielefeld, Germany\\
        E-mail: \email{sschlichting@physik.uni-bielefeld.de}}

\abstract{We calculate the evolution of a jet shower due to medium induced splittings in the deep LPM regime. Due to the characteristic energy dependence of the formation time $t_{\rm form}(\omega)= \sqrt{\omega /\hat{\bar{q}}}$, the radiative break-up process exhibits turbulent characteristics, allowing for analytic predictions of various inclusive properties of the medium induced cascade in an inertial range of energies $T \ll \omega \ll E$, where $E$ is the energy of the jet and $T$ is the temperature of the medium.}

\FullConference{International Conference on Hard and Electromagnetic Probes of High-Energy Nuclear Collisions\\
		30 September - 5 October 2018\\
		Aix-Les-Bains, Savoie, France}

\begin{document}

\section{Introduction}
Highly energetic particles or jets produced in collider experiments provide a unique tool to study the dynamics of strong-interaction matter. The evolution of jets in vacuum is theoretically well established within pQCD and serves as a precision tool in pp collisions; conversely the evolution of jets in a de-confined Quark-Gluon Plasma (QGP), created in the collision of heavy nuclei, can serve as a probe of the QGP medium and remains an active area of research. While a full characterization of jets in a QCD medium represents a formidable task, involving many different scales from below the the temperature of the medium $T \lesssim 1~{\rm GeV}$ to the energy of the jet $E \gtrsim 100~{\rm GeV}$, we will not attempt to provide a complete description of the dynamics. Instead, we will focus on one particular aspect of jet evolution in a QCD medium, namely the radiative break-up or fragmentation of a jet due to medium induced radiation~\cite{Mehtar-Tani:2018zba}. 

\section{Evolution of $\qua/\glu$ jets in the medium}
We describe the jet fragments as a collection of highly energetic ($\omega \gg T$) on-shell quarks/gluons propagating in a thermal QGP. Various kinds of elastic $(2\leftrightarrow 2)$ and inelastic $(1\leftrightarrow 2)$ interactions with the constituents of the thermal QGP can modify the properties of the jet~\cite{Ghiglieri:2015ala}. However, for highly energetic fragments $(\omega \gg T)$ the evolution is dominated by radiative branchings, where due to the interaction with the medium a hard parton of energy $\omega$, splits into two fragments with energy fractions $z\omega$ and $(1-z)\omega$. In the regime where in-medium splittings are induced by multiple soft interactions, as characterized by a momentum diffusion constant $\hat{\bar{q}}$,\footnote{Note that color factors are removed from $\hat{\bar{q}}$ and absorbed into the splitting functions instead.}  the corresponding splitting rates are given by~\cite{Ghiglieri:2015ala,Arnold:2008zu}
\begin{eqnarray}
\Gamma^{\rm split}_{fi}\left(\omega | z\omega,(1-z)\omega\right)= \frac{\alpha}{\pi} \frac{K_{fi}(z)}{t_{\rm form}(\omega)}
\end{eqnarray}
where $i,f=\qua,\glu$ labels the initial and final states and $K_{fi}(z)$ denote the in-medium splitting functions (see \cite{Mehtar-Tani:2018zba} for details). Due to the characteristic energy dependence of the formation time $t_{\rm form}(\omega)=\sqrt{\omega / \hat{\bar{q}}}$ associated with the Landau-Pomeranchuk-Migdal (LPM) effect \cite{Baier:1998kq,Zakharov:1996fv}, the splitting rates increase as $\Gamma^{\rm split}_{fi}(\omega) = \sqrt{E/\omega}~\Gamma^{\rm split}_{fi}(E)$ for less energetic particles, indicating the importance of successive splittings in the process. By treating successive splittings as quasi-instantaneous and independent of each other, the evolution of the in-medium fragmentation functions $D_{i}(x)\equiv \omega~dN_{i}/d\omega$, which describe the inclusive distribution of $i=\qua_f,\bar{\qua}_f,\glu$ fragments as function of the energy fractions $x=\omega/E$, is then described by a coupled set of effective kinetic equations~\cite{Mehtar-Tani:2018zba}
\begin{eqnarray}
\label{eq:Dg}
\partial_{\tau} D_{\glu}(x)=\int_{0}^{1} dz~\left\{\frac{K_{\glu\glu}(z) }{\sqrt{x/z}} D_{\glu}\left( \frac{x}{z} \right) + \frac{K_{\glu\qua}(z) }{\sqrt{x/z}} D_{S}\left( \frac{x}{z} \right) - \frac{K_{\glu\glu}(z) + K_{\qua\glu}(z)  }{2\sqrt{x}}  D_{\glu}(x)\right\}\;, \\
\label{eq:Dq}
\partial_{\tau} D_{S}(x)=\int_{0}^{1} dz~\left\{\frac{K_{\qua\glu}(z) }{\sqrt{x/z}} D_{\glu}\left( \frac{x}{z} \right) + \frac{K_{\qua\qua}(z) }{\sqrt{x/z}} D_{S}\left( \frac{x}{z} \right) - \frac{K_{\glu\qua}(z) + K_{\qua\qua}(z)  }{2\sqrt{x}}  D_{S}(x)\right\}\;,
\end{eqnarray}
for the gluon distribution $D_{\glu}$ and the flavor singlet quark distribution $D_{\rm S}=\sum_{f=1}^{N_f}(D_{q_f}+D_{\bar{q}_f})$.\footnote{We note that there is an additional set of kinetic equations for the evolution of the flavor non-singlet (valence) distributions $D_{\rm NS}^{(f)}=D_{q_f}-D_{\bar{q}_f}$, which decouples from Eqns.~(\ref{eq:Dg},\ref{eq:Dq}), and refer to \cite{Mehtar-Tani:2018zba} for the evolution of the valence distribution.}  Note that in Eqns.~(\ref{eq:Dg},\ref{eq:Dq}), the dependence on the medium properties $(\hat{\bar{q}})$ and the jet energy $(E)$ have been adsorbed into the definition of the characteristic time scale $\tau=\frac{\alpha}{\pi}~t/t_{\rm form}(E)$.

\begin{figure}[t!]
\begin{center}
\includegraphics[width=\textwidth]{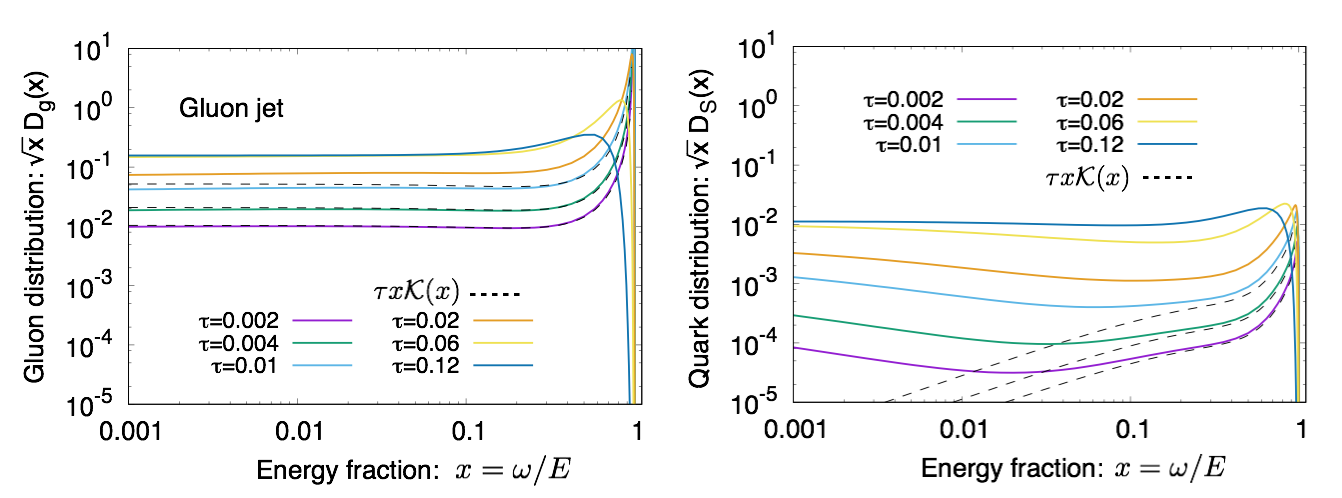}
\caption{\label{fig:gJet} Evolution of the in-medium fragmentation functions for gluons (left) and quarks (right) inside a gluon jet. Different curves in each panel show the results after evolution for different amounts of time $\tau$ inside the medium.  \cite{Mehtar-Tani:2018zba}}
\end{center}
\end{figure}

\subsection{Single vs. multiple splittings, wave turbulence \& universal quark/gluon ratio}

\begin{figure}[t!]
\begin{center}
\includegraphics[width=\textwidth]{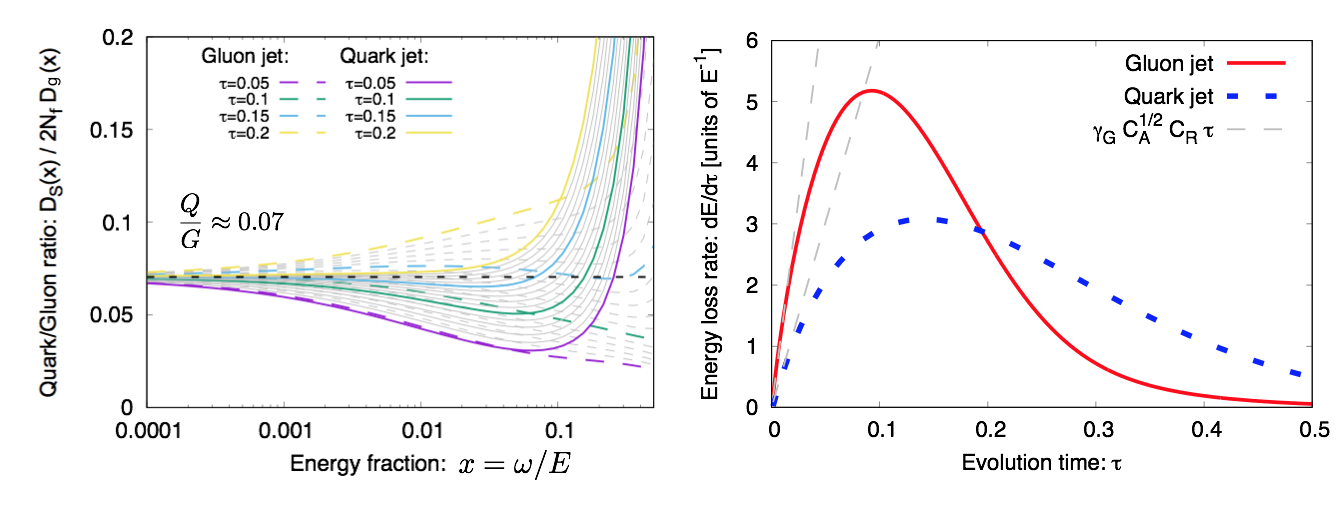}
\caption{\label{fig:two} (left) Quark/gluon ratio of jet fragments as a function of the energy fraction $x=\omega/E$. Different solid (dashed) curves correspond to results for quark (gluon) jets after evolution for different amounts of time $\tau$ inside the medium. The universal Kolmogorov ratio in Eq.~(\ref{eq:qgratio}) is indicated by the black dashed line. (right) Energy loss rate $dE/d\tau$ for quark and gluon jets. Dashed curves show the analytic estimate in Eq.~(\ref{eq:eloss}) for $\tau\ll1$.  }
\end{center}
\end{figure}

Numerical solutions of Eqns.~(\ref{eq:Dg},\ref{eq:Dq}) are presented in Fig.~\ref{fig:gJet}, which shows  fragmentation functions $\sqrt{x}~D_{\glu/S}(x,\tau)$ for a gluon jet ($D_{\glu}(x,\tau=0)=\delta(1-x)\;, D_{S}(x,\tau=0)=0$) after a short period of evolution $\tau = 0.002 - 0.12$ inside the medium. While at moderate values of $x$ the in-medium fragmentation functions are dominated by a single splitting, i.e. $D_{i}(x)\simeq\tau~x~K_{i\glu}(x)$, clear deviations from this behavior can be observed for smaller momentum fractions $x\lesssim 0.1$ already after very a short period of evolution inside the medium. Since the emission rates at small $x$ are enhanced by a factor $\Gamma^{\rm split}_{fi}(xE) = \frac{1}{\sqrt{x}}~\Gamma^{\rm split}_{fi}(E)$ due to the shorter formation times for softer radiation, the fragmentation process for $x\lesssim 0.1$ is dominated by multiple quasi-democratic splittings and proceeds in analogy to the famous Richardson cascade. In particular one finds that the in-medium fragmentation functions approach turbulent spectra of the Kolmogorov-Zakharov (KZ) form \cite{Baier:2000sb,Blaizot:2013hx,Mehtar-Tani:2018zba}
\begin{eqnarray}
D_{\glu}(x)=G/\sqrt{x}\;, \qquad D_{S}(x)=2N_fQ/\sqrt{x}\;,
\end{eqnarray}
associated with a scale $(x)$ independent energy flux
\begin{eqnarray}
\frac{1}{E} \frac{dE}{d\tau}=-\gamma_{\glu} G - \gamma_{\qua} Q\;, \qquad \gamma_{i}=\int_{0}^{1}dz\sum_{f=\qua,\glu} K_{fi}(z)~z \log(1/z)
\end{eqnarray}
directed towards smaller energies. A careful analysis of the stationary turbulent solutions of the kinetic equations~(\ref{eq:Dg},\ref{eq:Dq}) reveals that the chemistry of small $x$ fragments is uniquely fixed by the balance of $\glu\to\qua \bar{\qua}$ and $\qua\to\glu\qua$ processes, yielding~\cite{Mehtar-Tani:2018zba}
\begin{eqnarray}
\label{eq:qgratio}
\left.\frac{D_{S}(x)}{2N_f D_{g}(x)} \right|_{x\ll1}=\frac{Q}{G}= \frac{\int_{0}^{1}dz~z~K_{\qua\glu}(z)}{2N_f \int_{0}^{1}dz~z~K_{\glu \qua}(z)} \approx 0.07.
\end{eqnarray}
Numerical solutions of the evolution equations shown in the left panel of Fig.~\ref{fig:two} confirm that the universal ratio in Eq.~(\ref{eq:qgratio}), is indeed realized over a substantial range of momentum fractions $x$ and evolution times $\tau$ inside the medium, regardless of the nature ($\qua/\glu$) of the leading jet parton.

\subsection{Energy loss of quark \& gluon jets}
Besides affecting the jet substructure, the medium induced cascade also leads to an efficient energy transfer to the thermal medium, as shown in the right panel of Fig.~\ref{fig:two}, where we
present the rate $dE/d\tau$ of energy transferred to the thermal medium. Notably, for a short amount of evolution inside the medium $(\tau \ll 1)$ the energy loss rate can be understood as a simple two-step process~\cite{Baier:2000sb,Mehtar-Tani:2018zba}, where initially a single emission off the original hard parton creates a soft gluon spectrum of the form
$D_{g}(x,\tau) \simeq \tau~x~K_{\glu i}(x) \stackrel{x\ll1}{\simeq} \tau~C_{A}^{1/2} C_{R}/\sqrt{x}$. Subsequently, multiple splittings of the soft jet fragments create a turbulent energy flux
\begin{eqnarray}
\label{eq:eloss}
\frac{dE}{dt}=\frac{\alpha^2}{\pi^2}\hat{\bar{q}}~\gamma_{\glu} C_{A}^{1/2} C_{R}~t
\end{eqnarray}
transporting energy all the way to the scales of the medium $(x\sim T/E \ll 1)$. Since for $\tau \ll 1$, the turbulent cascade is initiated by the original hard parton, the energy loss in Eq.~(\ref{eq:eloss}) features the familiar Casimir scaling, such that $\left.dE/dt \right.|_{\rm q-jet} = C_F/C_A ~\left.dE/dt \right.|_{\rm g-jet} $. However, beyond time scales $\tau \gtrsim 0.1$ the chemical composition of the hard components of the jet is strongly altered by medium induced branchings, leading to a break down of the naive Casimir scaling seen in Fig.~\ref{fig:two}.

\section{Discussion}
Based on an effective kinetic description of jet evolution in a thermal medium, we find that the in-medium fragmentation of jets is governed by a turbulent cascade associated with the transport of energy from the scale of jet $(\omega \sim E)$ all the way to the scale of the medium $(\omega \sim T)$. Due to the turbulent nature of the process, important features of the medium induced cascade can be directly inferred from the stationary turbulent solutions of the underlying kinetic equations. Some of our our findings, such as the modified chemical composition of jet fragments, could also have interesting consequences for various jet sub-structure observables, including e.g. modified strangeness/heavy-flavor production in heavy-ion jets~\cite{Mehtar-Tani:2018zba}. However,  to further explore phenomenological consequences it will be necessary to combine the results of our study of the in-medium evolution with a theoretical description of the initial jet production and the hadronization of jet fragments. While this work is in progress, it is also worth pointing out that these phenomena could also be studied within Monte-Carlo event generators such as \textsc{MARTINI} or \textsc{JEWEL}, which are based on the same underlying microscopic processes.

\end{document}